\documentclass[structabstract]{aa}  
\usepackage{graphicx}
\usepackage{txfonts}
\usepackage{natbib}
\usepackage{multirow}

\begin{document}
   \title{The origin and orbit of the old, metal-rich, open cluster NGC~6791}

   \subtitle{Insights from  kinematics}

   \author{L.~J\'{i}lkov\'{a}\inst{1}\fnmsep\inst{2}
           \and
          G. Carraro\inst{1}\fnmsep\inst{3}
           \and
          B. Jungwiert\inst{4}\fnmsep\inst{5}
           \and
          I. Minchev\inst{6}
         }

   \institute{ESO, Alonso de Cordova 3107, Casilla 19001, Santiago, Chile
              \email{ljilkova,gcarraro@eso.org}
           \and
              Department of Theoretical Physics and Astrophysics, Faculty of Science, Masaryk University, Kotl\'a\v{r}sk\'a 2, CZ-611\,37 Brno, Czech Republic
           \and
              Dipartimento di Astronomia, Universit\'a di Padova, Vicolo Osservatorio 3, I-35122, Padova, Italy
           \and
              Astronomical Institute, Academy of Sciences of the Czech Republic, Bo\v{c}n\'{i} II 1401/1a, CZ-141 31 Prague, Czech Republic
            \email{bruno@ig.cas.cz}
           \and
              Astronomical Institute, Faculty of Mathematics and Physics, Charles University in Prague, Ke~Karlovu 3, CZ-121 16 Prague, Czech Republic
           \and 
              Leibniz-Institut f\"{u}r Astrophysik Potsdam (AIP), An der Sternwarte 16, D-14482 Potsdam, Germany
            \email{iminchev@aip.de}
             }

   \date{Received May 25, 2011 / accepted March 1, 2012}

% \abstract{}{}{}{}{} 
% 5 {} token are mandatory

%%%%%%%%%%%%%%%%%%%%%%%%%%%%%%%%%%%%%%%%%%%%%%%%%%%%%%%%%%%%%%%%% 
  \abstract
   {\object{NGC\,6791} is a unique stellar system among Galactic open clusters, which is at the same time one of the oldest open clusters and the most metal rich. It is located inside the solar circle, harbors a large population of binary stars, and possibly experienced prolonged star formation. The combination of all these properties is puzzling and poses the intriguing question of its origin.}
   {One possible scenario is that the cluster formed close to the Galactic Center and later migrated outward to its current location. In this work we study the cluster's orbit and investigate the possible migration processes that may have displaced NGC\,6791 to its present-day position, under the assumption that it actually formed in the inner disk.}
   {To this aim we performed integrations of NGC\,6791's orbit in a potential consistent with the main Milky Way parameters. In addition to analytical expressions for halo, bulge and disk, we also consider the effect of bar and spiral arm perturbations, which are expected to be very important for the disk dynamical evolution, especially inside the solar circle. Starting from state-of-the art initial conditions for NGC\,6791, we calculated 1000 orbits back in time for about 1\,Gyr turning different non-axisymmetric components of the global potential on and off. We then compared statistical estimates of the cluster's recent orbital parameters with the orbital parameters of $10^{4}$ test-particles originating close to the Galactic Center (with initial galocentric radii in the range of 3--5\,kpc) and undergoing radial migration during 8\,Gyr of forward integration.}
   {We find that a model that incorporates a strong bar and spiral arm perturbations can indeed be responsible for the migration of NGC\,6791 from the inner disk (galocentric radii of 3--5\,kpc) to its present-day location. Such a model can provide orbital parameters that are close enough to the observed ones. However, the probability of this scenario as it results from our investigations is very low.}
   {}

   \maketitle

%%%%%%%%%%%%%%%%%%%%%%%%%%%%%%%%%%%%%%%%%%%%%%%%%%%%%%%%%%%%%%%%%

\section{Introduction} \label{sec:intro}

Star clusters\,--\,globular and open\,--\,have long been recognized as ideal benchmarks for stellar evolution theories primarily because they are made of a large number of stars that share the same age, distance, and metallicity. This allows comparisons with stellar models in a statistical sense. The only exception has routinely been $\omega$\,Cen, which has been known for quite some time to display a significant spread in color in its red giant branch (RGB)  and inhomogeneities in chemical elements (\citealp{piotto05}, \citealp{villanova10}, and references therein).

However, in the past decades, signatures of multiple stellar populations \,--\,with different age and metallicity\,--\,have been detected in virtually all Milky Way (MW) globulars by means of very accurate photometry and spectroscopy (see \citealp{piotto09} for a review on the subject). Typical photometric signatures are the presence of multiple distinct main sequences (MS) or RGBs in the same cluster color magnitude diagram (CMD), while spectroscopic signatures are normally in the form of spreads in the abundance of iron and other elements much larger than observational uncertainties \citep{milone09}.

The rich, intermediate-age star clusters of the Magellanic Clouds (MCs) do not seem to be an exception \citep{milone09}, although in this case signatures of prolonged star formation are more common than evidences of discrete episodes of star formation. Indeed, in the CMD of MCs clusters the MS is much broader than expected by photometric errors and binary stars, but does not separate into discrete sequences unlike in Galactic globulars.

As of now, the only MW open cluster where evidence of prolonged star formation has been found is NGC\,6791\,. The  upper MS and turn-off region of this cluster are much wider than expected from photometric errors and binary stars alone. By separating the stars lying inside the cluster core and outside it, \citet{twarog11} found that inner stars occupy the red edge of the upper MS, while outer stars occupy the blue edge. The most viable explanation for this occurrence is an age spread of $\sim 1$\,Gyr between the inner and the outer regions of the cluster, in the sense that inner stars formed before outer stars. Star formation therefore started probably in the center of the cluster and continued outward during quite a long time.

NGC\,6791 is unique in the MW because it is one of the most massive old open clusters with an age of $\sim$\,8\,Gyr, and is extremely metal-rich ([Fe/H] $\sim +0.40$, \citealp{carraro06}, \citealp{origlia06}). No other open cluster in the MW is that rich in metals. This combination of age and metallicity is unique in the whole population of Galactic open clusters. While the RGB of NGC\,6791 is broad in color \citep{janes84}, no statistically significant evidence for a metallicity spread within the cluster\,--\,which would support a self-enrichment scenario\,--\,has been reported so far. Within observational errors, stars in NGC\,6791 show the same metal abundance. As mentioned by \citet{twarog11}, {\it since the CMD-based age spread estimated for populous clusters is usually less than the 0.7\,Gyr found for NGC\,419 by \citet{rubele10}, the CMD of NGC\,6791 may have resembled that of the Magellanic Cloud clusters 7\,Gyr ago.}

The cluster's current location is at 4\,kpc from the Sun, close to the solar circle, at 8\,kpc from the Galactic Center (GC) and about 0.8\,kpc above the plane \citep[derived from ][]{brogaard11}. The metallicity of this cluster is at odds with the radial abundance gradient defined by the other MW open clusters \citep{carraro06}.

This wealth of unique properties clearly poses the basic question of the origin of this system. So far, to answer to this question only arguments based on the observed chemical characteristics of the cluster have been used. If this system is so metal rich, it must have formed very close to the Galactic bulge and then been deposited where it is now by some migration process. Indeed, only in the inner regions of the MW can star formation and metal enrichment be fast and strong enough to generate a~stellar system like NGC\,6791.

Alternatively, one can argue that NGC\,6791 is the left-over of a dwarf elliptical that was disrupted by the MW some time ago and was then engulfed by the Galaxy, losing most of its mass by tidal interaction. This seems to be supported by the many extreme blue horizontal branch stars discovered in the cluster \citep{kaluzny95,liebert94,buson06} and by the recent measure of a significant UV-upturn in the spectral energy distribution of the cluster \citep{buzzoni11}. Such a UV-upturn is only found in moderately UV-enhanced ellipticals and has the same strength as in, e.g., NGC\,4552 and NGC\,4486, which emit $\sim$ 2$\%$ of their light at a~shorter wavelength than 2500 \AA. This feature does not have any counterpart in the population of Galactic open clusters.

The study of the NGC\,6791 Galactic orbit did not reveal much to clarify the scenario so far, since depending on the quality of present-day kinematic data, simple backward calculation could produce different orbits\,--\,a regular orbit with different eccentricities $e\sim$\,0.5 or $\sim$\,0.3 by \citet{bedin06}, and \citet{wu09}, respectively, both reaching galocentric radii of $\sim$\,10\,kpc, and an orbit by \citet{carraro06} extending to more than 20\,kpc in the Galactic disk.

In this paper we present a numerical study of the NGC\,6791 Galactic orbit that includes the combined effect of the Galactic bar and spiral arms into an analytical MW model. At the same time,  we aim at investigating in detail the scenario in which NGC\,6791 formed close to the GC, where star formation might have been rapid and efficient enough to generate it. This scenario requires some  radial migration process to move the cluster's orbit outward in the Galactic disk.
 
Several different dynamical mechanisms triggering radial mixing are discussed in the recent literature:

\begin{itemize}
\item[1.] the resonant scattering by transient spiral structure was firstly introduced by \citet{sellwood02}. Stars can move inward and outward at the corotation radius while preserving their circular orbit. During the evolution of the disk the angular velocity (the corotation radius) changes and the spirals can induce radial migration within the whole disk. The importance of the transient spiral structure radial migration phenomenon was confirmed by \citet{roskar08} \citep[see also][]{roskar11}. 

\item[2.] \citet{minchev10a} described another diffusion mechanism resulting from the resonance overlap of multiple rotating patterns, the Galactic bar, and spiral arms. Using test-particle simulations, these authors found that the overlap mechanism can be up to an order of magnitude more effective than the one due to transient spiral structure (for the model consistent with the MW it takes $\sim 3$\,Gyr to produce the same radial mixing as in $\sim 9$\,Gyr by the transients, see also \citealp{minchev10c}). This migration mechanism is a strong function of the strength of MW bar and spiral arms. However, this dependency is non-linear. The characteristics of the resonance overlap mechanism were later confirmed by \citet{minchev11} via a fully self-consistent, tree-SPH N-body simulations. Among other results, these authors also found that the  presence of a gaseous component makes the migration even more efficient.

\item[3.] Small satellites have been found to effectively mix the outer disks of galaxies. \citet{quillen09} used disk-like distributed test-particles perturbed by a low-mass satellite (of mass a few times $10^9\,$M$_{\sun}$). These authors find that the disk stars from the outer disk can move to low eccentricity orbits inside their birth radii. \citet{bird11} compared radial migration in several fully self-consistent N-body simulations both with and without interaction with satellites. Among other conclusions, these authors showed that the radial migration induced by satellites is a distinct mechanism that does not happen in isolated systems.
\end{itemize}

During the past decades it has become widely accepted that the MW has a central bar. Evidence for this comes from observations of the inner Galaxy (e.g., NIR photometry\,--\,\citealt{blitz91}, \citealt{dwek95}, \citealt{binney97}; gas kinematics\,--\,\citealt{englmaier99}; and microlensing\,--\,\citealt{rattenbury07}) and dynamical effects near the solar circle \citep{dehnen00,minchev07,minchev10d}. It is, therefore, very important to consider the MW's bar in orbital integration studies, especially in the inner disk, but also near the Sun.

While for the \citet{sellwood02} mechanism to work short-lived transient spirals are required, in barred galaxies, such as the MW, spirals are more likely coupled with the bar, and are thus longer-lived \citep{galdyn08,quillen10}. Therefore, in this paper we will concentrate on the migration induced by spiral-bar interaction, rather than transient spirals. The migration induced by orbiting satellites will also be unimportant for the inner disk.

In Sect.~\ref{sec:galmod} we describe the adopted analytical Galactic model and introduce spiral arm and bar perturbation. The initial conditions and the results of the backward orbit integration are presented in Sect.~\ref{sec:orbit}. We concentrate on radial migration in Sect.~\ref{sec:mig} and, finally, discuss the results of this work in Sect.~\ref{sec:concl}.

%%%%%%%%%%%%%%%%%%%%%%%%%%%%%%%%%%%%%%%%%%%%%%%%%%%%%%%%%%%%%%%%%

\section{Galactic models and orbit integrator} \label{sec:galmod}

To compute any Galactic orbit, it is necessary to adopt a model for the  MW gravitational potential. The potential should reproduce the observed mass density of the MW. While the  general picture of the today's MW structure is more or less well established and supported by observations, a more detailed description is still under lively discussion, see e.g., \citet{schonrich10} or \citet{coskun10} for the recent analysis of the velocity of the local standard of rest (LSR), or \citet{reid09} for the up-to-date description of MW rotation curve. These questions naturally bring difficultly quantifiable uncertainties for the construction of any MW model.

We chose to use a relatively simple model composed of several components. The axisymmetric time-independent part was modeled by a bulge, disk and halo. Additionally,  we included a~non-axisymmetric component\,--\,bar and spiral arms\,--\,that can act separately or together. For each component, an analytic expression for the density is  provided, from which the corresponding potential and force can be derived.

The model is described in the following subsection. The main idea of our study is to qualitatively estimate the viability of the migration induced by the bar and spiral arms resonance overlap mechanism relative to the migration in simpler potentials, either purely axisymmetric or containing just one type of non-axisymmetric perturbation. 

\subsection{Axisymmetric model} \label{sec:as}
For the time-independent axi\-sym\-met\-ric component of the Galactic potential we used the model introduced by \citet{flynn96} that was also used by \citet{gardner10}. 
\footnote{We also tested the model of \citet{allen91}, slightly modified to follow the more recent observations rotation curve as described below. The nature of the general results presented in Sects.~\ref{sec:orbit} and \ref{sec:mig} stays similar as in the case when the axisymmetric component is modeled as described in Sect.~\ref{sec:as}. In its simplicity the model introduced by \citet{allen91} does not follow the most updated knowledge of the MW mass distribution. However, it has been recently and extensively used by several teams dealing with orbits of MW clusters \citep[e.g., ][]{allen06,allen08,bellini10,wu09} or with the solar neighborhood kinematics \citep{antoja09,antoja11}.}
This model has three components\,--\,bulge, disk, and dark halo. The bulge is modeled as a superposition of two Plummer spheres \citep{plummer11}. The disk is modeled as a superposition of three Miyamoto---Nagai disks \citep{miyamoto75}. Finally, the dark halo is modeled as a spherical logarithmic potential. For clarity, we give the equation for each component of the axisymmetric potential:
\begin{eqnarray}
\phi_0 &=& \phi_{\mathrm{C}} + \phi_{\mathrm{D}} + \phi_{\mathrm{H}}, \\
\phi_{\mathrm{C}} &=& - \frac{GM_{\mathrm{C_1}}}{\sqrt{r^2+r_{\mathrm{C_1}}^2}} - \frac{GM_{\mathrm{C_2}}}{\sqrt{r^2+r_{\mathrm{C_2}}^2}}, \\
\phi_{\mathrm{D}_n} &=& \frac{-GM_{\mathrm{D}_n}}{\sqrt{\left\{R^2 + \left[a_n + \sqrt{(z^2 + b^2)}\right]^2\right\}}},\quad n=1, 2, 3, \\
\phi_{\mathrm{H}} &=& \frac{1}{2} V_{\mathrm{H}}^2 \ln\left(r^2 + r_0^2\right),
\end{eqnarray}
where $r$ is the galocentric radius ($r^2 = R^2 + z^2$), $G$ is the gravitational constant, $M_{\mathrm{C_1}}$ and $M_{\mathrm{C_2}}$,  $r_{\mathrm{C_1}}$ and $r_{\mathrm{C_2}}$, are the masses and scale radii of the bulge components, respectively. $M_{\mathrm{D}_n}$ and $a_n$ are masses and scale-lengths of the three disk components ($n$ goes from 1 to 3). The parameter $b$ is related to the scale-height of the disk and is the same for all three components of the disk. The halo has core radius $r_0$ and  $V_{\mathrm{H}}$ is the asymptotic circular velocity at large radii (relative to $r_0$).

\citet{flynn96} chose the parameters of their model to reproduce the current MW observations at that time. We modified the values of some parameters to achieve a model consistent with more recent MW observations. As for the recent observations of the rotation curve at the solar galocentric distance $R_{\sun}$, \citet{reid09} estimated $R_{\sun} = (8.4 \pm 0.6)$\,kpc and the solar circular rotation speed $\Theta_{\sun} = (254 \pm 16)$\,km\,s$^{-1}$. To derive these values, they used the solar motion with respect to the LSR, $v_{\sun,\mathrm{LSR}}$, as determined  by \citet{dehnen98} from the Hipparcos stars. The same Hipparcos data set was later re-examined by \citet{schonrich10}, who obtained slightly different values for the $v_{\sun,\mathrm{LSR}}$ components. Namely, they found the velocity component in the direction of the Galactic rotation to be about 7\,km\,s$^{-1}$ higher than \citet{dehnen98}. Considering these more recent results, we changed the original \citet{flynn96} values for the halo circular velocity $V_{\mathrm{H}}$ and the disk parameters (masses and scale radii) to obtain $\Theta_{\sun} = 243$\,km\,s$^{-1}$ at $R_{\sun} = 8.4$\,kpc. These values are consistent with the $\Theta_{\sun}$ derived from the apparent proper motion of Sgr\,A* \citep{reid04} using $R_{\sun} = 8.4$\,kpc and the \citet{schonrich10} solar motion.

The superposition of three Miyamoto---Nagai disks builds up a disk with exponential falloff of the radial surface density on a wide range of radii. Similar as in \citet{gardner10}, we chose the disk parameters to obtain the scale-length of the exponential falloff of the disk consistent with recent observations. Our disk gives a radial surface density profile approximately corresponding to the exponential falloff with the scale-length of 2.6\,kpc for radii of 5--18\,kpc, which is consistent with recent observations \citep{juric08}.

Values for all parameters of the axisymmetrical component are given in Table~\ref{tab:galmod}.

\begin{table}
\centering
\caption{Parameters of the axisymmetrical component of today's MW model.}
\label{tab:galmod}
\begin{tabular}{r l c}
\hline\hline
Component & Parameter & Value \\
\hline
Bulge & $M_{\mathrm{C_1}}$ & 0.3$\cdot 10^{10}$M$_{\sun}$ \\
      & $r_{\mathrm{C_1}}$ & 2.7\,kpc \\
      & $M_{\mathrm{C_2}}$ & 1.6$\cdot 10^{10}$M$_{\sun}$ \\
      & $r_{\mathrm{C_2}}$ & 0.42\,kpc \\
Disk  & $M_{\mathrm{D}_1}$ & 8.9$\cdot 10^{10}$M$_{\sun}$ \\
      & $a_1$              & 5.0\,kpc \\
      & $M_{\mathrm{D}_2}$ & -6.9$\cdot 10^{10}$M$_{\sun}$ \\
      & $a_2$              & 15.8\,kpc \\
      & $M_{\mathrm{D}_3}$ & 2.8$\cdot 10^{10}$M$_{\sun}$ \\
      & $a_3$              & 33.0\,kpc \\
      & $b$                & 0.3\,kpc \\
Halo  & $V_{\mathrm{H}}$   & 225 km\,s$^{-1}$\\
      & $r_0$              & 8.4\,kpc\\
\hline
\end{tabular}
\end{table}

The above described axisymmetric potential was also used in the models including the non-axisymmetric components. However, when we included the bar perturbation, the mass $M_{\mathrm{C_2}}$ of one of the bulge components was reduced and the bar replaced most of the bulge mass. We kept the axisymmetric components unaltered when we turn the spiral arms perturbation on (see Subsection \ref{subsec:sa}).

\subsection{Galactic bar}
For the bar model we adopted the Ferrers potential \citep[see paragraph 2.5.3 (a) of][ and we chose $n=2$]{galdyn08} of an inhomogeneous triaxial ellipsoid \citep{pfenniger84}. The bar model is defined by  the bar length, axes ratios, mass, and the angular velocity. For all parameters (listed in Table~\ref{tab:galmod_na}) we adopted the same values as reported by \citet{pichardo04}\, and refer the reader to their original paper for a justification of the various choices.

\subsection{Spiral arms}\label{subsec:sa}
The parameters of the spiral structure are much more uncertain than for the bar. Most observational studies suggest two-armed or four-armed spiral pattern, or their super-position (see for example \citealt{vallee05}; \citealt{vallee08}; \citealt{allen08}; \citealt{grosbol11}).

We modeled the gravitational potential perturbation induced by spiral arms as described  by \citet{cox02}. We used a two-armed spiral pattern with their basic sinusoidal arms. The perturbation has the shape of a logarithmic spiral. The shape of the spiral arms model is defined by five parameters\,--\,radial scale length, scale height, pitch angle, amplitude of density perturbation (mass density of the arms in the Galactic plane at the solar radius), and angular velocity. The vertical structure of the arms' mass density has the form  $\mathrm{sech}^{2}(z/z_{\mathrm{s}})$, where $z_{\mathrm{s}}$ is a scale height. We chose the values of parameters consistent with a current picture of the MW spiral structure\,--\,see Table~\ref{tab:galmod_na}. 

\citet{drimmel01} present a three-dimensional model for the MW fitted to the COBE far-infrared and near-infrared emission. They found that two-armed structure dominates the near-infrared emission and the arm-interarm density contrast of their modeled stellar spiral arms has a value of about 1.8 at the solar radius \citep[see Fig.~15 in][]{drimmel01}. Our choice of parameters for \citet{cox02} gives 2.0 for the same ratio. The value is also consistent with the observations of external galaxies with a weaker spiral structure \citep[see, e.g.,][]{block04}.

In addition to the parameters defining the shape of spiral arms, we needed to specify the initial orientation of the arms. We followed the result of the study of the COBE/DIRBE $K$-band emission made by \citet{drimmel00}, which is consistent with two-armed spiral model of the MW. 

\begin{table}
\centering
\caption{Parameters of the non-axisymmetrical components in today's MW model (indicated as MW1 in Sect.~\ref{sec:mig}) with their references.}
\label{tab:galmod_na}
\begin{tabular}{c c c}
\hline\hline
Parameter & Value & Reference \\
\hline
\multicolumn{3}{c}{Galactic bar} \\
%\hline
Bar mass                & 0.98$\cdot 10^{10}$M$_{\sun}$ & 
\multirow{6}{*}
{$\left. \begin{array}{c} \\ \\ \\ \\ \\ \\ \end{array} \right\}$  (1)} \\
Bulge mass $M_{\mathrm{C_2}}$ & 0.62$\cdot 10^{9}$M$_{\sun}$ & \\
Angular velocity    & 60.0\,km\,s$^{-1}$\,kpc$^{-1}$ & \\
Major axis half-length & 3.14\,kpc & \\
Axial ratios        & 10:3.75:2.56 & \\
Bar angle & 20$\degr$ & \\
\hline
\multicolumn{3}{c}{Spiral arms} \\
%\hline
Number of spiral arms     & 2                            & \\
Pitch angle      & 15.5$\degr$                  & (2) \\
Amplitude        & 3.36\,$\cdot 10^{7}$M$_{\sun}$\,kpc$^{-3}$ & \\
Angular velocity & 20.0\,km\,s$^{-1}$\,kpc$^{-1}$ & \\
Scale-length     & 2.6\,kpc              & (3)\\
Scale-height     & 0.3\,kpc              & (3) \\
Scale-radius     & 5.6\,kpc              & (2) \\
\hline
\end{tabular}
\tablebib{(1) \citet{pichardo04}; (2) \citet{drimmel00}; 
(3) \citet{juric08}}
\end{table}

\begin{figure}
  \centering
  \includegraphics[width=8.7cm]{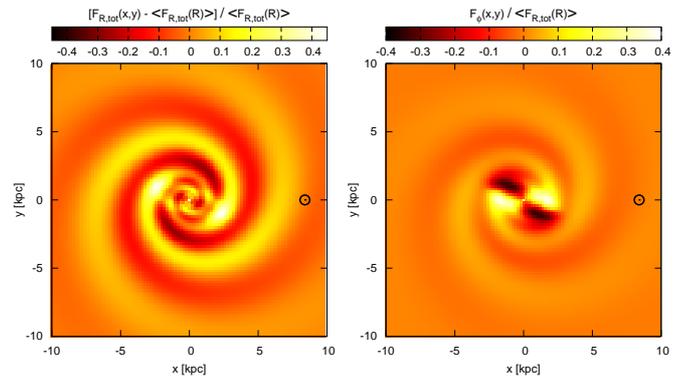}
  \caption{Maps of strength of non-axisymmetric components in the combined (bar and spiral arms) today's MW model (noted as MW1 in Sect.~\ref{sec:mig}). Color scales map the ratio between radial (left) and tangential (right) force resulting from the non-monopole part of the gravitational potential to the radial force due to the monopole part at the same radius. Maps are plotted for the Galactic plane. $\sun$ symbol marks the position of the Sun.}
  \label{fig:fmap}
\end{figure}

\subsection{Strength of rotating patterns} \label{sec:strenght}
We estimated the contribution of the non-axisymmetric components by computing the force resulting from the bar or/and spiral arms' gra\-vi\-tational potentials in the disk plane, following the approach of \citet{combes81} \citep[see also][]{buta01}. Fig.~\ref{fig:fmap} shows maps of the ratio between the radial, resp. the tangential, force due to the non-monopole part of the combined (bar and spiral arms) potential to the radial force due to the monopole part at the same radius.

The strength of the bar is often characterized by the bar strength parameter, which is defined as the maximal value of the ratio between the bar's tangential force to the monopole radial force at the same radius. Our choice of bar model and parameters gives a bar strength parameter of 0.4. The model of spiral arms gives a value of 0.05 (using the tangential force from the spiral arms instead of the bar in the previous definition). 

\subsection{Orbit calculation}
Once a model of gravitational potential is adopted, the corresponding equations of motion can be constructed. We used Hamilton's equations, which give a set of first-order differential equations that we integrated using a Bulirsch-Stoer integrator with adaptive time-steps \citep{press92}. The relative change in the Jacobi constant for the models with a single rotating pattern is estimated to be on the order of $10^{-10}$.

To see and understand the different orbits caused by the bar or/and spiral arms we calculated orbits in four versions of the MW model\,--\,axisymmetric potential, potential with the bar but not spiral arms, with spiral arms and no bar and, finally, a combined model with both bar and spiral arms.

%%%%%%%%%%%%%%%%%%%%%%%%%%%%%%%%%%%%%%%%%%%%%%%%%%%%%%%%%%%%%%%%%

\section{Recent orbit from backward integration} \label{sec:orbit}

To investigate the shape and properties of the recent orbit of NGC\,6791, we integrated its orbit backward for  1\,Gyr, starting from the cluster's current position and velocity vectors. These initial vectors were obtained from the most recent observational estimates we found in the literature: equatorial celestial coordinates ($\alpha$ and  $\delta$ for J2000.0 equinox), distance to the Sun $d_{\sun}$, radial velocity $v_r$, and PM.

\begin{table}
\centering
\caption{Input observational data used for the calculation of the initial conditions\,--\,equatorial coordinates $\alpha$, $\delta$ (J2000), radial velocity $v_r$, PM $\mu_{\alpha} \cos \delta$, $\mu_{\delta}$, and distance to the Sun $d_{\odot}$; together with their references.}
\label{tab:ini}
\begin{tabular}{c r@{$\pm$}l c}
\hline\hline
 & \multicolumn{2}{c}{Adopted value} & Reference \\
\hline
 $\alpha$                  & \multicolumn{2}{c}{$290.22083\degr$} & (1) \\
 $\delta$                  & \multicolumn{2}{c}{$37.77167\degr$}  & (1) \\
 $v_r$                     &  $-47.1$  & 0.7\,km\,s$^{-1}$ & (2) \\ 
 $\mu_{\alpha} \cos \delta$ &  $-0.57$  & 0.13\,mas\,yr$^{-1}$ & (2) \\ 
 $\mu_{\delta}$             &  $-2.45$  & 0.12\,mas\,yr$^{-1}$ & (2) \\
 $d_{\odot}$                &   4.01    & 0.14\,kpc  & (3) \\
\hline
\end{tabular}
\tablebib{(1)~WEBDA; (2)~\citet{bedin06} (see also references there-in); (3)~\citet{brogaard11}.}
\end{table}

\subsection{Initial conditions} \label{sec:ini}
As mentioned above, NGC~6791 has been the subject of intense study in the past. However, we rely in this study on the most recent values for the aforementioned parameters, which we list in Table~\ref{tab:ini} together with their source. So that they can be used as initial conditions for orbit integration, the input quantities reported in Table~\ref{tab:ini} need to be transformed into the appropriate coordinate system. 

We integrated the orbits in a cartesian galocentric right-handed coordinates system: the $x$ axis points from the GC outward, in the direction of the Sun; the $y$ axis points in anti-direction of the LSR motion; and the $z$ axis toward the North Galactic Pole. The position of the Sun is given by vector $(x,y,z)_{\sun} = (8.4,0,0)$\,kpc in this coordinate system. The transformation of the observational data to the cartesian coordinate system centered on the Sun was performed following step by step the \citet{johnson87} algorithm updated to the International Celestial Reference System (the coordinates of the North Galactic Pole $\alpha=192.85948\degr$, $\delta=27.12825\degr$). Adopting the solar motion with respect to the LSR ($v_{\sun,\mathrm{LSR}}$) from \citet{schonrich10}: $(U,V,W)_{\sun} = (11.1,12.24,7.25)$\,km\,s$^{-1}$ (right-handed system, with $U$ in the direction to the GC and $V$ in the Galactic rotation direction)\,--\,the velocity vector with respect to LSR was obtained. Finally, using the Sun galocentric distance of 8.4\,kpc and the LSR rotation velocity of 243\,km\,s$^{-1}$ (see Sect.~\ref{sec:as}), the position and velocity vectors were transformed into the coordinate system of integrations (denoted as $x$, $y$, $z$, $v_x$, $v_y$, $v_z$).

\begin{figure*}
  \sidecaption
  \includegraphics[width=12cm]{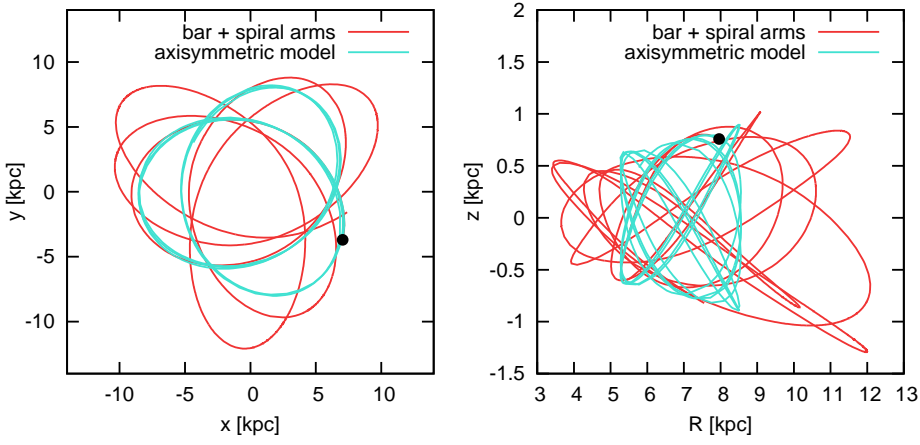}
  \caption{Recent portion of Galactic orbit of NGC\,6791\,--\,orbit projections into the Galactic and meridional plane are plotted in the left and right column, respectively. Orbit in axisymmetric potential is plotted with the blue line. The red line shows the combined model (bar and spiral arms). The initial conditions (see Table~\ref{tab:inc}) are given by mean observational input data (Table\,\ref{tab:ini}), the black dot marks the initial position. Integration was backward for time of 1\,Gyr.}
  \label{fig:orb}
\end{figure*}

\begin{figure*}
  \centering
  \includegraphics[width=17cm]{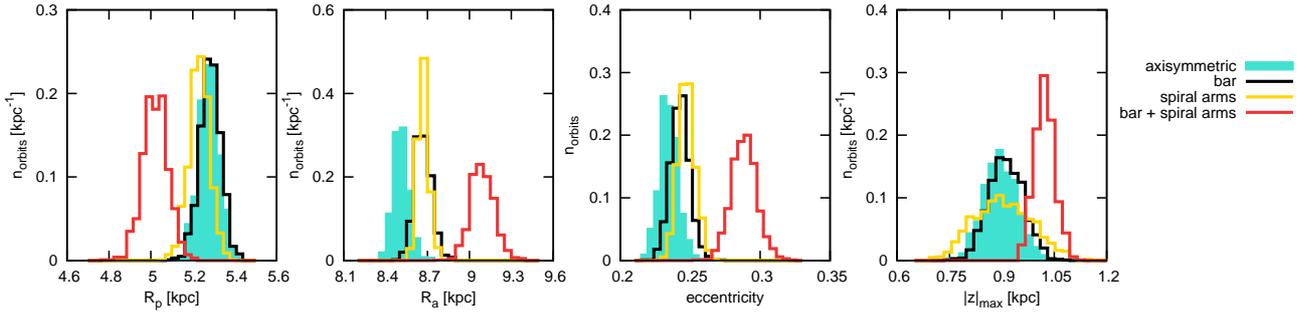}
  \caption{Distributions of cluster's recent orbital parameters. The distributions for peri-galacticon $R_{\mathrm{p}}$, apo-galacticon $R_{\mathrm{a}}$, vertical height of orbit $|z|_{\mathrm{max}}$, and eccentricity $e$ are plotted in columns. Different colors show results for different MW models: axisymmetric model\,--\,light blue, model with bar\,--\,black, model with spiral arms\,--\,yellow, and red for the bar and spiral arms combined model. }
  \label{fig:opar}
\end{figure*}

The input observational data come naturally with some associated uncertainties, which translate into uncertainties in the initial condition vectors. To take the observational uncertainties into account, we followed \citet{dinescu99} \citep[see also][]{wu09,vandeputte10}, and calculated a set of 1000 initial condition realizations. The initial velocity and position vectors were extracted from normally distributed values of radial velocity, PM, and distance to the Sun. The dispersions of a given distribution was taken  as the observational uncertainty of the input data from Table~\ref{tab:ini}. We integrated an orbit for each initial condition vector backward for 1\,Gyr. The initial positions and velocities ($x$, $y$, $z$, $v_x$, $v_y$, $v_z$) derived from the mean observational input data are given in Table~\ref{tab:inc}.

\begin{table*}
  \centering
  \caption{Initial conditions for orbit calculations derived from the mean values of observational input data (Table~\ref{tab:ini}), together with standard deviations of their distributions.}
  \label{tab:inc}
  \begin{tabular}{cccccccccccc}
  \hline\hline
   $x$ & $\sigma_{x}$ & $y$ & $\sigma_{y}$ & $z$ & $\sigma_{z}$ & $v_x$ &  $\sigma_{v_{x}}$ & $v_y$ & $\sigma_{v_{y}}$ & $v_z$  & $\sigma_{v_{z}}$ \\

   \multicolumn{2}{c}{[kpc]} & \multicolumn{2}{c}{[kpc]} & \multicolumn{2}{c}{[kpc]} & \multicolumn{2}{c}{[km\,s$^{-1}$]} & \multicolumn{2}{c}{[km\,s$^{-1}$]} & \multicolumn{2}{c}{[km\,s$^{-1}$]} \\
  \hline
   7.05 & 0.04 & -3.70 & 0.13 & 0.76 & 0.03 & -39.7 & 2.7 & -197.7 & 1.3 & -12.1 & 2.6 \\
  \hline
  \end{tabular}
  \tablefoot{See Sect.~\ref{sec:ini} for the description of the used coordinate system.}
\end{table*}

\subsection{Orbits and orbital parameters} \label{sec:orbpar}
Galactic orbits are routinely characterized by  a set of so-called orbital parameters: the peri-galacticon $R_{\mathrm{p}}$, the apo-galacticon $R_{\mathrm{a}}$, the vertical amplitude $|z|_{\mathrm{max}}$ defined as the maximal distance from the Galactic plane and, finally, the  eccentricity $e$ defined as $(R_{\mathrm{a}} - R_{\mathrm{p}})/(R_{\mathrm{a}} + R_{\mathrm{p}})$. Depending on the type of orbit, values of orbital parameters can differ for individual revolutions.

For irregular orbits orbital parameters change with time as the orbit evolves. This is particularly the case if the orbit is near a resonance for a single pattern or near a resonance overlap for bar$+$spiral arms. Since chaotic behavior of orbits is expected in these regions, it is impossible to recover/predict an orbit on longer timescales. Therefore, to see the current orbital history, we calculated the orbital parameters for the most recent revolution (defined by azimuthal change of 2$\pi$) for each set of initial conditions. The distributions of these orbital parameters are shown in Fig.~\ref{fig:opar}. 

To highlight and quantify the influence of the various non-axisymmetric components, we integrated orbits in all four MW models described in Sect.~\ref{sec:galmod}. The orbits (backward integration for time of 1\,Gyr) in axisymmetric and combined models given by the mean values of observational input data (see Table~\ref{tab:inc} for corresponding initial conditions) are plotted in Fig.\,\ref{fig:orb}\,--\,projection into the Galactic and meridional plane on the left and right, respectively. 

The orbits in models including only bar or spiral arms differ only slightly (around $10\%$) from the results of the axisymmetric model (see distributions of orbital parameters in Fig.~\ref{fig:opar}). With increasing model complexity there is a trend of having larger apo-galacticon and lower peri-galacticon, which, in turn, yields a higher eccentricity. This is never higher than 0.33, however, which seems to make highly eccentric orbits for NGC\,6791 implausible.
 
According to these calculations, NGC\,6791 is not expected to move much outside the solar circle, at odds with previous suggestions (see next Sect.~\ref{sec:compar}). The orbit is presently located close to its peri-galacticon and this remains inside the solar circle on average, but never gets very close to the bulge. However, a~scenario in which NGC\,6791 formed close to the bulge and then migrated outward deserves a closer look, which we shall present in Sect.~\ref{sec:mig}.

\subsection{Comparison with previous orbit integration} \label{sec:compar}
To our knowledge, the orbit of NGC\,6791 was studied three times\,--\,by \citet{carraro06}, \citet{bedin06}, and by \citet{wu09}.

\citet{carraro06} used complex potential derived from an N-body gas-dynamical model \citep{fux97,fux99}. They obtained an eccentric orbit ($e = 0.59$) extending beyond the solar circle. However, the initial conditions were derived from absolute PM and radial velocity values available at that moment and admittedly less precise than the ones we are using here, 
and therefore it cannot be compared directly with the results presented here.

\citet{bedin06} used a purely axisymmetric \citet{allen91} model with the most up-to-date PM and radial velocity (the same values as we use here). However, reviewing the details of the initial conditions set-up, we identified a problem (mis-definition of the $x$~axis orientation), which prevents a~proper comparison with our results.

\citet{wu09} calculated the orbit using the \citet{allen91} and two other axisymmetric potential models. As \citet{carraro06}, they derived the initial conditions from less precise data than we use here, but still their orbital parameters for the \citet{allen91} potential are within error-bars consistent with results given in this study.

In any case, the present study supersedes any previous one, in the sense that statistical analysis is performed for the orbital parameters adopting different Galactic model versions.

%%%%%%%%%%%%%%%%%%%%%%%%%%%%%%%%%%%%%%%%%%%%%%%%%%%%%%%%%%%%%%%%%

\section{Forward orbit integration and radial migration}
\label{sec:mig}

\begin{table}
\centering
\caption{Parameters of initial condition distributions for the forward orbits integration. We used cylindrical galocentric coordinates\,--\,galocentric radius $R$, azimuth $\phi$, vertical coordinate $z$ and corresponding velocities $v_{R}$, $v_{\phi}$, $v_{z}$. The tangential velocity $v_{\varphi}$ was calculated as the circular velocity $v_{\mathrm{c}}(R)$ in the axisymmetric potential as described in Sect.~\ref{sec:as} at a given radius and then a random velocity with a dispersion of 6\,km\,s$^{-1}$ was added.}
\label{tab:dist}
\begin{tabular}{cccc}
\hline\hline
Coordinate & Distribution & Center & Dispersion, width\\
\hline
$R$    & uniform & 4.0\,kpc & 1.0\,kpc \\
$\phi$ & \multicolumn{3}{c}{0--2$\pi$ with equidistant distribution} \\
$z$    & normal  & 0\,kpc  & 0.05\,kpc \\
$v_{R}$ & normal & 0\,km\,s$^{-1}$ & 10\,km\,s$^{-1}$ \\
$v_{\phi}$ & normal & $v_{\mathrm{c}}(R)$ & 6\,km\,s$^{-1}$ \\
$v_{z}$ & normal & 0\,km\,s$^{-1}$ & 4\,km\,s$^{-1}$ \\
\hline
\end{tabular}
\end{table}

\begin{figure*}
  \centering
  \includegraphics[width=17cm]{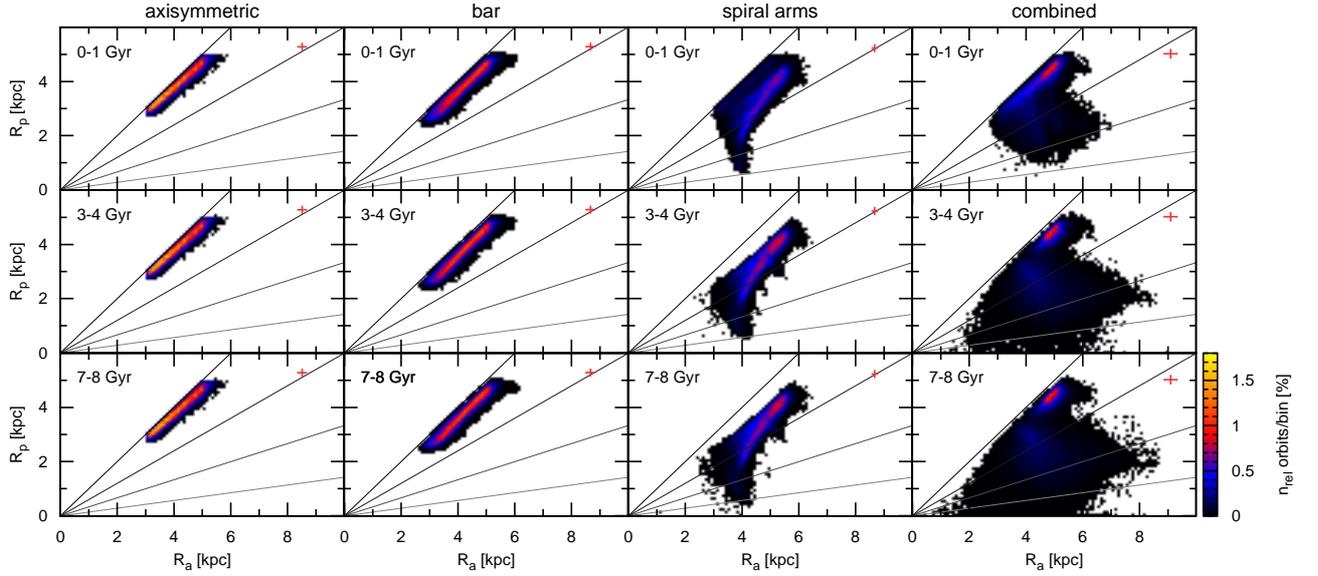}
  \caption{Time evolution of the apo-galacticon $R_{\mathrm{a}}$ against peri-galacticon $R_{\mathrm{p}}$ plane for the MW1 model. A mean probability of finding an orbit within~a given bin of $R_{\mathrm{a}}$ and $R_{\mathrm{p}}$ (where the mean is calculated over the indicated 1\,Gyr time interval) is mapped. The bin size is 0.1$\times$0.1\,kpc. The probability for every bin is calculated as a sum of orbits with the bin's $R_{\mathrm{a}}$ and $R_{\mathrm{p}}$ values, where the contribution of each orbit is divided by the total number of revolutions per orbit within the time interval (so the total contribution of each orbit to each map is equal to 1). Finally, the probability is normalized to the total number of orbits $10^4$ and converted to \%. Each column shows orbits for models including different components\,--\,axisymmetric model, model with bar, model with spiral arms, and finally bar and spiral arms combined model. Orbits are mapped within time intervals of 1\,Gyr (0--1, 3--4, and 7--8\,Gyr) as indicated in the left upper corner of each plot. The NGC\,6791's values of $R_{\mathrm{a}}$ and $R_{\mathrm{p}}$ with their three sigma error-bars are plotted with a red cross in each plot. The four lines in each of the plots show the lines of constant eccentricity for values 0.0 (circular orbits), 0.25, 0.5, and 0.75, respectively, with decreasing slope.}
  \label{fig:ap_mw1}
\end{figure*}

\begin{figure*}
  \centering
  \includegraphics[width=17cm]{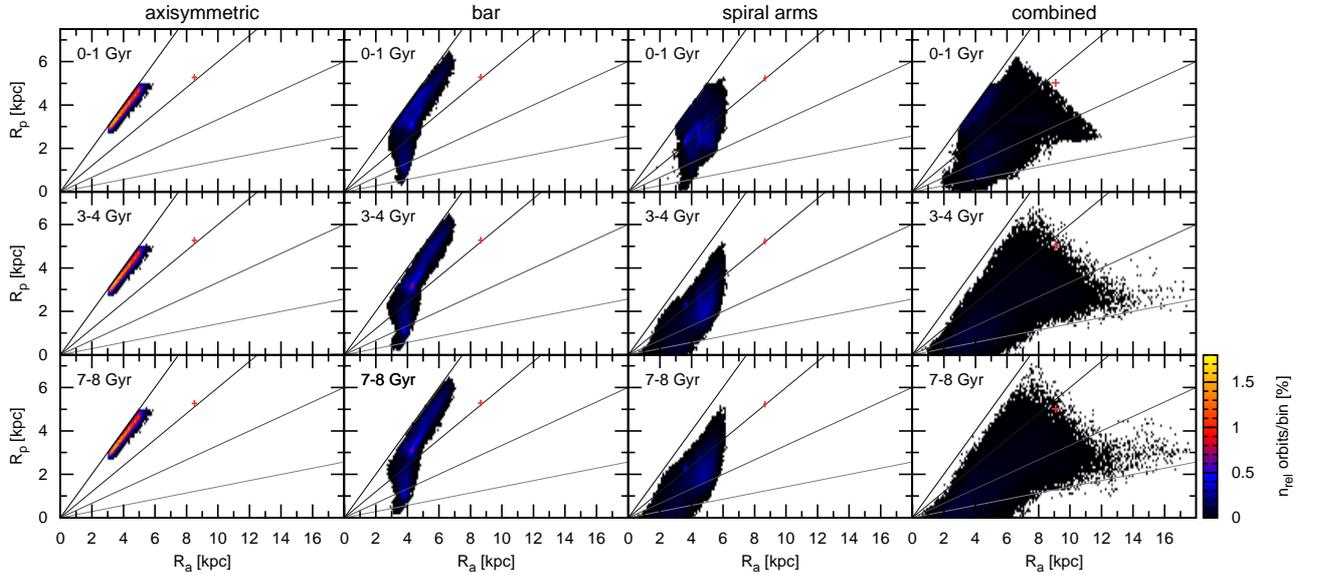}
  \caption{Time evolution of the apo-galacticon $R_{\mathrm{a}}$ against peri-galacticon $R_{\mathrm{p}}$ plane for the MW2 model, see Fig.~\ref{fig:ap_mw1} for a detailed description of the plot.}
  \label{fig:ap_mw2}
\end{figure*}

\begin{figure*}
  \centering
  \includegraphics[width=17cm]{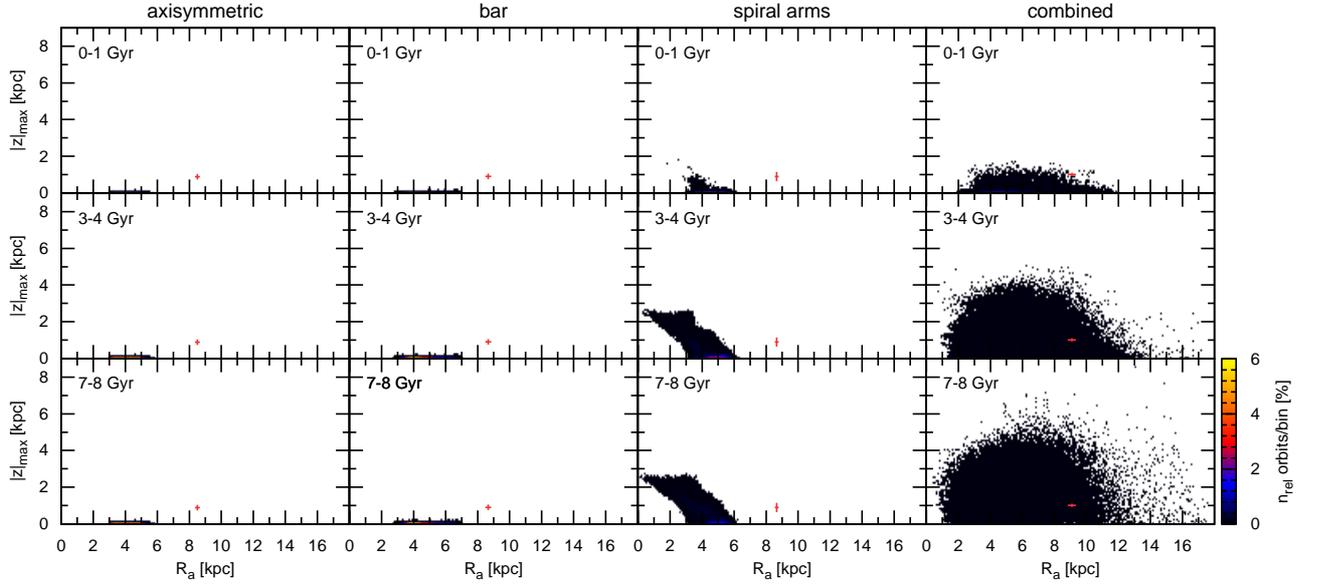}
  \caption{Time evolution of the apo-galacticon $R_{\mathrm{a}}$ against vertical amplitude $|z|_{\mathrm{max}}$ plane for the MW2 model. Similar to Fig.~\ref{fig:ap_mw1}, results for the models including different components are plotted in columns; time increases with rows.}
  \label{fig:az_mw2}
\end{figure*}

\begin{figure*}
  \centering
  \includegraphics[width=17cm]{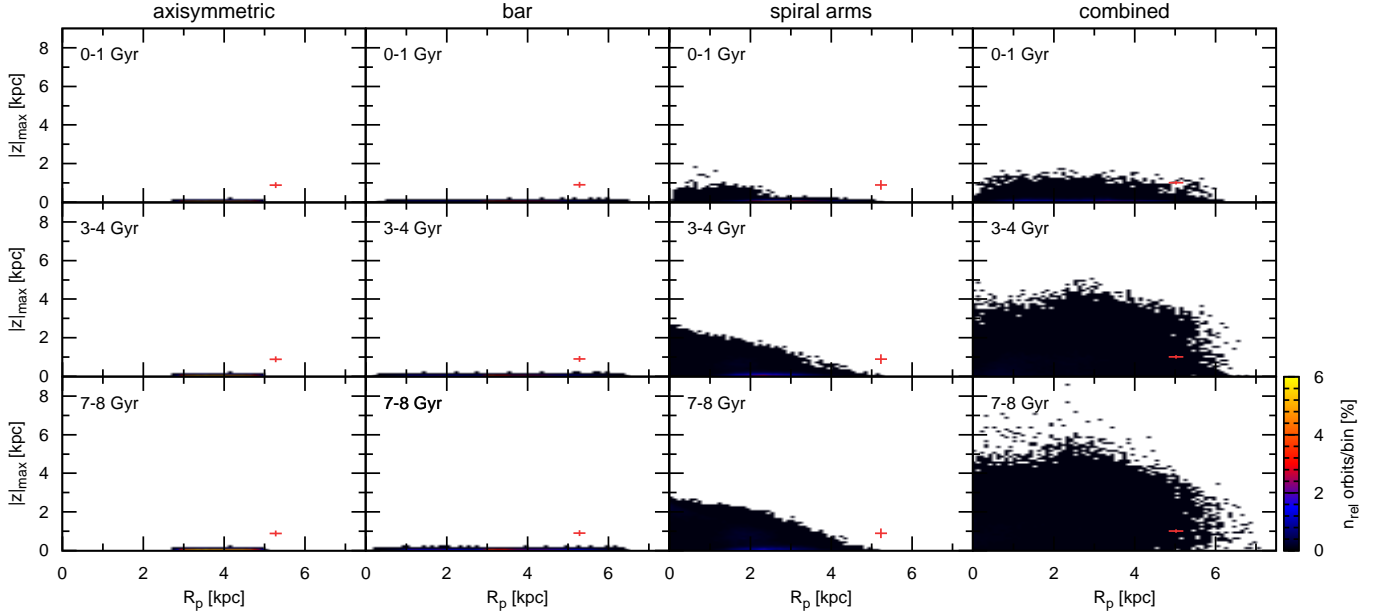}
  \caption{Time evolution of the peri-galacticon $R_{\mathrm{p}}$ against vertical amplitude $|z|_{\mathrm{max}}$ plane for the MW2 model. See Fig.~\ref{fig:ap_mw1} for a detailed description.}
  \label{fig:pz_mw2}
\end{figure*}

\begin{table}
 \centering
 \caption{Parameters for MW models used for the forward integrations.}
 \label{tab:mwmod}
 \begin{tabular}{cccc}
 \hline\hline
 \multicolumn{2}{c}{Parameter} & MW1 & MW2 \\
 \hline
 Bar angular velocity & [km\,s$^{-1}$\,kpc$^{-1}$] & 60.0 & 45.0 \\
 Bulge mass $M_{\mathrm{C_2}}$ & [$10^{10}$M$_{\sun}$] & 0.62  & 0   \\
 Bar mass   & [$10^{10}$M$_{\sun}$] & 0.98  & 1.6 \\
 Major axis & [kpc] & 3.14 & 5.0 \\
 spiral arms' angular velocity & [km\,s$^{-1}$\,kpc$^{-1}$] & 20 & 30 \\
 spiral arms' amplitude & [$10^{7}$M$_{\sun}$\,kpc$^{-3}$] & 3.36 & 5.14 \\
 spiral arms' pitch angle & [$\deg$] & 15.5 & 20.0 \\                 
 \hline
 \end{tabular}
 \tablefoot{Both models have the same form and are characterized by the same parameters as the model used for forward integrations described in Sect.~\ref{sec:galmod} (denoted as model MW1 here). In model MW2, listed parameters are changed (the rest of model parameters are kept the same as for MW1 model, see Table~\ref{tab:galmod_na}).}
\end{table}

The inner disk, close to the bulge, is a high-density region where star formation is very efficient and metal enrichment fast \citep{bensby10}. It is therefore conceivable to imagine that NGC~6791 could have formed there and then diffused outward. To estimate the probability and efficiency of radial migration we carried out a forward orbits integration. More specifically, we integrated forward in time a number of orbits with an initial location close to the GC. By comparing their orbital properties with the recent NGC\,6791 orbital parameters obtained in Sect.~\ref{sec:orbit}, we attempted to estimate the probability that radial migration moved the NGC~6791 orbit outward from the inner Galactic regions.

We followed $10^4$ initially close-to-circular orbits distributed between 3 and 5\,kpc from the GC. Initial positions of orbits were equidistantly separated in azimuth. Then, the radius was generated randomly for each azimuth, from a uniform distribution between 3--5\,kpc (see Table~\ref{tab:dist} for a detailed description of initial condition distributions in phase-space). 

To quantify the relative influence of different non-axisymmetric components, we followed the orbits in all four flavors of the Galactic potential model described in Sect.~\ref{sec:galmod} (axisymmetric, barred, spiral, combined).  When included, the amplitude of the non-axisymmetric components grows simultaneously in time for 0.4\,Gyr from 0 to its maximum value with the time-dependency described by \citet[][given by his equation (4)]{dehnen00}. The gradual growth of non-axisymmetric perturbations is introduced to allow a progressive adaptation of initially nearly circular orbits. The growth of a real bar and real spiral arms is certainly much more complex and not simultaneous, but modeling these time dependencies is beyond the scope of our model. We tested that our results are not significantly sensitive, apart from the first few hundreds of Myr, to the above gradual growth when compared to orbit integrations in which the bar and/or spiral arms were switched on abruptly to their full strength from the beginning. 

According to \citet{minchev10a} 3\,Gyr should be sufficient for disk radial mixing. On the other hand, the age of the NGC\,6791 cluster is estimated to be $\sim$8\,Gyr as mentioned in Sect.~\ref{sec:intro}. Therefore, we integrated the orbits forward in time for 8\,Gyr. Finally, the orbital parameters were calculated for individual rotations of each orbit as in the Sect.~\ref{sec:orbpar}.

The properties of the bar and spiral arms that we used for the integrations of the NGC\,6791's orbit (as listed in Table~\ref{tab:galmod_na}) are based on recent observations of the MW (see Sect.~\ref{sec:galmod}). We indicate this present-day MW model as MW1. From N-body simulations and observations of external galaxies at various red-shifts, we can infer  the evolution of bar and spiral arms with time. For example, owing to the expected increase of stellar velocity dispersion with time, the disk becomes more stable and thus its spiral structure weakens. On the other hand, gas inflow can cool down the disk and the structure can strengthen again. Central mass concentrations (e.g., gas inflow) can decrease bar amplitudes \citep[e.g.,][]{athanassoula05,debattista06} and even destroy bars, reforming later from the newly formed stars, but with a decreased strength and size \citep[e.g.,][]{bournaud02,combes08}. 

Therefore, although we do not know the MW bar and spiral structure history, we can safely assume that those were much stronger in the past. For this reason we also followed orbits in a~potential including stronger and faster spiral arms, as well as a~stronger bar. This model is denoted as MW2. In Table~\ref{tab:mwmod} we list the values of the parameters for which MW1 and MW2 differ. We decided to show an example of a model that is deliberately tuned to favor the radial migration caused by stronger bar and spiral arms and also by the choice of the rotating speed of the spiral arms pattern, bringing its resonances closer to the GC. We note that the parameter space (bar and spiral arms shape, extent, strength, and pattern speed) within the framework of our test particle integrations in analytically expressed potentials is very large and its systematic exploration was not the goal of the paper.

\begin{table}
 \centering
 \caption{Radii (in kpc) of the most important resonances for the Galactic bar and spiral arms MW models\,--\,corotation $R_{\mathrm{CR}}$, inner Lindblad resonances $R_{\mathrm{ILR}}$, and outer Lindblad resonances $R_{\mathrm{OLR}}$.}
 \label{tab:res}
 \begin{tabular}{ccccc}
 \hline\hline
 \multirow{2}{*}{Resonance} &  \multicolumn{2}{c}{MW1} & \multicolumn{2}{c}{MW2} \\
 & bar & spiral arms & bar & spiral arms \\
 \hline
 $R_{\mathrm{CR}}$  & 3.6 & 12.3 & 5.1 & 8.1 \\
 $R_{\mathrm{ILR}}$ & 1.4 & 2.8  & 1.7 & 2.2 \\
 $R_{\mathrm{OLR}}$ & 6.9 & 20.4 & 9.4 & 13.9 \\
 \hline
 \end{tabular}
 \tablefoot{Values for both non-axisymmetric components and both MW models are given. See Tables~\ref{tab:galmod_na} and~\ref{tab:mwmod} for angular velocities and other model characteristics. The axisymmetric model described in Sect.~\ref{sec:as} is used for the unperturbed potential.}
\end{table}

\begin{figure}
  \centering
  \includegraphics[width=8.7cm]{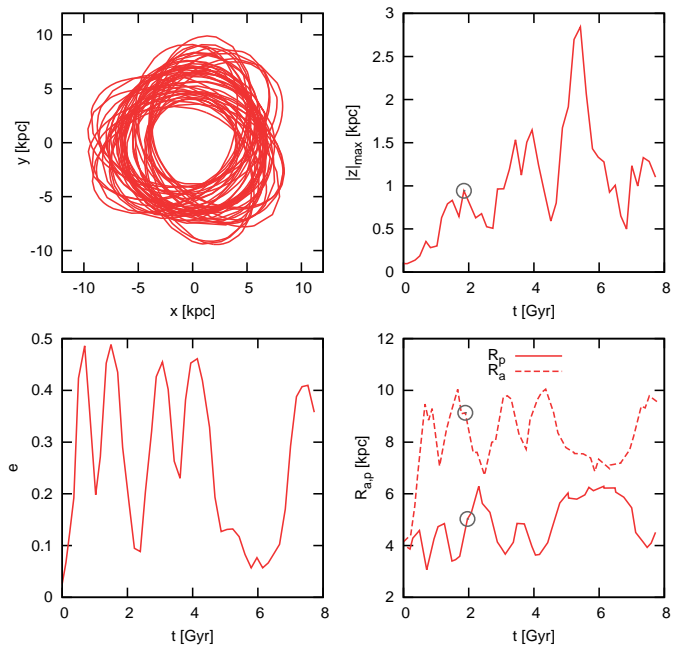}
  \caption{One of the Galactic orbits and its orbital parameters evolutions calculated in the MW2 model including bar and spiral arms\,--\,upper left plot shows the orbit in the Galactic plane; upper right plot shows the time-evolution of the vertical amplitude; the bottom row shows evolutions of eccentricity, and peri- and apo-galacticon in the left and right plot, respectively. For one of its revolutions, this orbit fits the 3$\sigma$ intervals for peri-galacticon, apo-galacticon, and vertical amplitude of the NGC\,6791 orbit, these moments are marked with circle symbols in plots on the right.}
  \label{fig:mig_orb}
\end{figure}

\begin{figure}
  \centering
  \includegraphics[width=8.7cm]{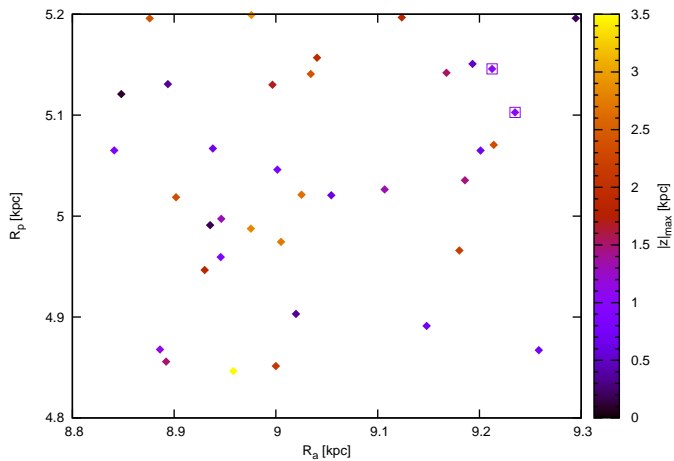}
  \caption{Apo-galacticon $R_{\mathrm{a}}$ against peri-galacticon $R_{\mathrm{p}}$ plane for orbits integrated in the MW2 model during the time interval of 7--8\,Gyr. The range of $R_{\mathrm{a}}$ as-well-as of $R_{\mathrm{p}}$ is limited to the $3\sigma$ intervals of orbital parameters values of NGC\,6791. Each revolution with this $R_{\mathrm{a}}$ and $R_{\mathrm{p}}$ is shown by a point. The color of each point maps the vertical amplitude $|z|_{\mathrm{max}}$. Revolutions that match NGC\,6791 for all three orbital parameters (within their $3\sigma$ intervals) are marked with squared boxes.}
  \label{fig:mw2_apz}
\end{figure}
 
In detail, the bar semi-major axis of the MW2 model was chosen to be longer, of 5\,kpc (but the axis ratio for other two axes was kept as for MW1 shape of bar) and the bar itself was more massive\,--\,so that on a time scale of 0.4\,Gyr it replaced the whole mass of the bulge. As a consequence, the strength parameter of this bar is 0.5. As for spiral arms, the spiral pattern was set to have a higher amplitude and higher pitch angle. The angular velocity of the bar in this case was decreased to 45\,km\,s$^{-1}$\,kpc$^{-1}$. This brings the bar's corotation radius to 5.1\,kpc and so the ratio of bar's corotation radius to its semi-major axis is consistent with the value of 0.9--1.3 found in most barred galaxies \citep[see paragraph 6.5.1 in][ and references therein]{galdyn08}. On the other hand, the spiral arms angular velocity is increased to 30\,km\,s$^{-1}$\,kpc$^{-1}$. The resonances of spiral arms are therefore closer to the GC where the orbits are originally distributed. See Table~\ref{tab:res} for resonance radii in both models. 

Before starting to comment more on the results, we recall the reader that we aimed to search for similarities between the present-day orbital parameters of NGC~6791 and the orbital parameters of orbits that started close to the GC and then migrated outward. If there were similarities, we could claim that the scenario in which NGC~6791 could form in the very inner disk is plausible also from a dynamical point of view, and not supported only by the observed chemical characteristics of the cluster.

The outcome of these experiments is illustrated in Figs.~\ref{fig:ap_mw1}\,--\,\ref{fig:mw2_apz}. Figs.~\ref{fig:ap_mw1} and \ref{fig:ap_mw2} map the evolution of mean probability of orbit with a given combination of $R_{\textrm{a}}$ and $R_{\textrm{p}}$. For the direct comparison with the recent orbital parameters of NGC\,6791 we also indicate its values with the red cross in the maps. Figs.~\ref{fig:az_mw2} and \ref{fig:pz_mw2} map the mean probability in the vertical amplitude $|z|_{\textrm{max}}$ against the $R_{\textrm{a}}$ and $R_{\textrm{p}}$ plane for the MW2 model, respectively. As mentioned above, model MW2 was chosen to obtain an enhanced migration. Comparing Figs.~\ref{fig:ap_mw1} and \ref{fig:ap_mw2}, we can see that this is indeed the case, illustrated by the higher values of apo-galacticon seen in Fig.~\ref{fig:ap_mw2}. Moreover, some orbits have rotations with similar orbital parameters as we derived for NGC\,6791 (the position of the red cross in the  $R_{\textrm{a}}$ against $R_{\textrm{p}}$ plane is within the area populated by some of the forward integrated orbits). 

From Figs.~\ref{fig:az_mw2} and \ref{fig:pz_mw2} we see that only the MW2 combined model can produce orbits that also show a vertical amplitude similar to NGC\,6791. Only very few orbits (169 out of $10^4$) can at least once per the total integration time of 8\,Gyr reproduce the $R_{\textrm{a}}$ and $R_{\textrm{p}}$ values similar to NGC\,6791, and only some of these orbits (13 out of the 169) meet all three orbital parameters together. To investigate these orbits in more detail, we display all three orbital parameters together in Fig.~\ref{fig:mw2_apz} for the time interval of 7--8\,Gyr of the integration. 

The experiment with forward integrated orbits illustrated in Figs.~\ref{fig:ap_mw1} to~\ref{fig:mw2_apz} helps us to provide the following considerations:

\begin{itemize}

\item[$\bullet$] only model MW2 that includes bar and spiral arms produces the apo-, peri-galacticons, and eventually also the vertical amplitude similar to the backward integration of the current orbit (see the maps for the combined model in the rightmost column of Figs.~\ref{fig:ap_mw2}--\ref{fig:pz_mw2}). 

\item[$\bullet$] Fig.~\ref{fig:mig_orb} shows an example of an orbit that migrated close to the recent NGC\,6791 one. Initially, the orbit is close to circular with the radius of 4\,kpc, and later starts to oscillate between very eccentric and close to circular again but has a radius of about 7\,kpc.

\item[$\bullet$] from our experiment, we estimate the probability for the MW2 model including bar and spiral arms to produce the present orbit of NGC\,6791 restricted to 2D Galactic plane to be about 0.40\% (during the total integration time of 8\,Gyr). This probability is calculated in the same way as for maps in Figs.~\ref{fig:ap_mw1}--\ref{fig:pz_mw2}: as a sum of orbits reaching $R_{\mathrm{a}}$ and $R_{\mathrm{p}}$ within $3\sigma$ intervals of NGC\,6791, where the contribution of each orbit is divided by the total number of revolutions per orbit within the time interval of 1\,Gyr (so the total contribution of each orbit within the time interval is equal to 1). Finally, the probability is normalized to the total number of orbits $10^4$ and we sum the probabilities for time intervals of 1\,Gyr within the whole integration time of 8\,Gyr. If the vertical amplitude is taken into account, the probability has an even lower value of 0.03\%.

\end{itemize}

%%%%%%%%%%%%%%%%%%%%%%%%%%%%%%%%%%%%%%%%%%%%%%%%%%%%%%%%%%%%%%%%%

\section{Summary and conclusions}
\label{sec:concl}

We investigated a scenario in which the old, metal-rich open cluster NGC\,6791 formed in the inner disk, close to the bulge, and then moved outward because of perturbations induced by the bar and spiral arms. This scenario might indeed explain its unique properties, since star formation close to the bulge is strong and metal enrichment fast. NGC\,6791's large mass and compactness can explain why it has survived to the present time.

We conducted our investigation using a numerical tool that allowed us to integrate orbits in different Galactic potentials. We calculated the recent (1\,Gyr back in time) Galactic orbit of NGC\,6791 in various MW models\,--\,axisymmetric,  including bar and/or spiral structure. The observational uncertainties were taken into account in Monte Carlo fashion, calculating a  set of 1000 orbits with their initial conditions distributed according to PM, radial velocity, and distance observational uncertainties. We found that orbits in the axisymmetric model and models including only one rotating pattern differ only slightly. The orbit in the combined bar and spiral arms model is more eccentric with eccentricity up to $\sim$0.3, and has a higher vertical amplitude of $\sim$1\,kpc.

To investigate whether these orbital parameters can be achieved as a consequence of migration processes from the inner disk, we also followed a set of forward integrations (for 8\,Gyr) with initial positions closer to the GC (with galocentric radii of 3--5\,kpc). To this aim, we considered two different realizations of the MW\,--\,present-day model: the same we used for the backward integrations (MW1), and models with rotating patterns more supporting the radial migration  process (MW2).

The purpose of this analysis was not to reproduce the cluster dynamical history precisely, but to estimate how efficient the migration induced by the MW bar and spiral structure could be and how high the probability of the orbit realization within the limitations of our model is. We indeed found that our MW2 model that incorporates both the bar and spiral arm perturbations, can produce orbits with apo-galacticon and peri-galacticon similar to the actual values of NGC\,6791. However, the probability of this scenario, as quantified from our experiment, is low, approximately 0.40\%. Moreover, our migration scenario struggles to reproduce the vertical amplitude of NGC\,6791. This might be an artifact of the \citet{flynn96} potential, which does not represent the vertical structure of the disk well over the studied range of galocentric distances, as well as of uncertainties in the modeled vertical structure of the bar and spiral arms.  We are aware that this could also imply a problem with the migration scenario itself, question that is unfortunately insolvable within the limitations of the current model.

Our study is unable to provide evidence for the inner origin of NGC\,6791 and its later displacement due to radial migration caused by the bar and the spiral arms resonance overlap. This could be viewed as an argument in support of the external (accreted dwarf satellite) origin of NGC 6791, in line with the UV upturn argument mentioned in Sect.~\ref{sec:intro}. On the other hand, our simple model gives a non-zero probability of migration from the inner galaxy, which would therefore deserve additional investigation with improved models of the MW and with a more systematically covering of model's parameter space.

The recent results by \citet{minchev10a} and \citet{minchev11} show that the addition of a second perturbation creates stochastic regions throughout the galactic disk and thus allows for a large-scale migration without resorting to the effect of short-lived transients. We note that for transient spirals this effect would be amplified. We expect that the addition of a secondary spiral density wave, as found in N-body simulations of barred systems \citep[][ Minchev et al., in preparation]{quillen10}, that may possibly exist in the MW \citep{naoz07}, will make the migration process more efficient. Therefore, the strengths of bar and spirals we used in this study may be the upper limits on what is required to push NGC\,6791 outward.
 
Future studies should strive to incorporate more realistic dynamical models, e.g., considering the time evolution of the bar and spiral structure parameters, the effect of a gaseous component initially present in the disk, and continuous gas accretion.

%%%%%%%%%%%%%%%%%%%%%%%%%%%%%%%%%%%%%%%%%%%%%%%%%%%%%%%%%%%%%%%%%

\begin{acknowledgements}
This research has made use of the WEBDA database, operated at the Institute for Astronomy of the University of Vienna (\texttt{http://www.univie.ac.at/webda/}). LJ acknowledges the support of two-year ESO PhD studentship, held in ESO, Santiago, as well as by grants No.\,205/08/H005 (Czech Science Foundation) and MUNI/A/0968/2009 (Masaryk University in Brno). BJ was supported by the grant LC06014-Center for Theoretical Astrophysics (Czech Ministry of Education) and by the research plan AV0Z10030501 (Academy of Sciences of the Czech Republic). We thank the anonymous referee, whose valuable comments and suggestions helped to significantly improve the paper.
\end{acknowledgements}

%%%%%%%%%%%%%%%%%%%%%%%%%%%%%%%%%%%%%%%%%%%%%%%%%%%%%%%%%%%%%%%%%

\bibliographystyle{aa}
\bibliography{paper}

%%%%%%%%%%%%%%%%%%%%%%%%%%%%%%%%%%%%%%%%%%%%%%%%%%%%%%%%%%%%%%%%%

\end{document}